\documentclass[11pt]{article} 
\usepackage{graphicx,floatflt,amssymb,epsf,rotate} 
\begin{document} 
\begin{flushright} 
HRI-P-09-09-001 \\
RECAPP-HRI-2009-017\\
CU-Physics/5 - 2009
\end{flushright} 
 
\vskip 30pt 
 
\begin{center} 
{\Large \bf GUTs with dim-5 interactions:}
\vskip 5pt 
{\Large \bf Gauge Unification and Intermediate Scales} \\
\vspace*{1cm} 
\renewcommand{\thefootnote}{\fnsymbol{footnote}} 
{ {\sf Joydeep
Chakrabortty${}^1$\footnote{e-mail: joydeep@hri.res.in}  and}
 {\sf Amitava Raychaudhuri${}^{1,2}$} 
} \\ 
\vspace{10pt} 
   ${}^{1)}$ {\em Harish-Chandra Research Institute,\\
Chhatnag Road, Jhunsi, Allahabad  211 019, India}  

   ${}^{2)}$ {\em Department of Physics, University of Calcutta, \\
92 Acharya Prafulla Chandra Road, Kolkata 700 009, India}  \\
\normalsize 
\end{center} 
 
\begin{abstract} 
\noindent
Dimension-5 corrections to the gauge kinetic term of Grand
Unified Theories (GUTs) may capture effects of quantum gravity or
string compactification. Such operators modify the usual gauge
coupling unification prediction in a calculable manner. Here we
examine SU(5), SO(10), and E(6) GUTs in the light of all such permitted
operators and calculate the impact on the intermediate scales and
the unification programme. We show that in many cases at least one
intermediate scale can be lowered to even 1-10 TeV, where a
neutral $Z'$ and possibly other states are expected.

\vskip 5pt 
\noindent 
\texttt{PACS Nos:~ 12.10.Dm, 12.10.Kt, 11.10.Hi } \\ 
\texttt{Key Words:~~Grand Unified Theories, ~Dimension-5 operator } 
\end{abstract}

\renewcommand{\thesection}{\Roman{section}} 
\setcounter{footnote}{0} 
\renewcommand{\thefootnote}{\arabic{footnote}} 
\noindent

\section{Introduction} 

Grand Unified Theories (GUTs)  \cite{guts} relate the strong and
electroweak interactions of the Standard Model (SM) at a high
energy, $M_X$, and embody quark-lepton unification, leading to
testable predictions such as proton decay and $n-\bar{n}$
oscillations. The characteristic energy of the SM, which is based
on the gauge group  ${\mathcal G}_{SM} \equiv SU(2)_L \otimes
U(1)_Y \otimes SU(3)_c $, is the electroweak scale $M_Z$.   The vast
difference between  $M_X$ and   $M_Z$ introduces a hierarchy
problem in GUTs which is often addressed through the introduction
of supersymmetry (SUSY). The rich predictions of these theories
-- of both the non-supersymmetric and supersymmetric varieties --
have received much attention. At the moment a clear experimental
confirmation of the GUT paradigm is keenly awaited.

The fourth fundamental interaction, namely, gravity, is not a
part of GUTs. It is widely expected that grand unified theories
will have a setting in some larger framework, e.g., string
theory, effective at higher energies close to the Planck scale,
$M_{Pl}$, which will encompass gravitational interactions within
its fold.  Without going into the details of such a theory one
can hope to probe some of its implications through effective
operators at the GUT scale, suppressed by inverse powers of
$M_{Pl}$,  which may emerge from it and alter the
grand unified theory predictions.

The particular higher dimensional operator which we consider here
impacts the gauge kinetic term:
\begin{equation}
\mathcal{L}_{kin}=-\frac{1}{4 c} Tr(F_{\mu\nu} F^{\mu\nu}) .
\label{eq:kin}
\end{equation}
where $F^{\mu\nu} = \Sigma_i \lambda_{i}.F_{i}^{\mu\nu}$ is the
gauge field strength tensor with $\lambda_i$ being the matrix
representations of the generators normalised to
$Tr(\lambda_{i}\lambda_{j})=c~\delta_{ij}$. For
$SU(n)$ groups the $\lambda_{i}$ are conventionally chosen in the
fundamental representation with $c = 1/2$. 
 
The dimension-5 (dim-5) interaction which we include is \cite{prev,cr}:
\begin{equation}
\mathcal{L}_{dim-5}=-\frac{\eta}{M_{Pl}}\left[\frac{1}{4
c}Tr(F_{\mu\nu}\Phi_{D} F^{\mu\nu})\right] ,
\label{eq:dim5op}
\end{equation}
where $\Phi_{D}$ denotes the $D$-component Higgs multiplet and
$\eta$ parametrises the strength of this interaction. In order
for it to be possible to form a gauge invariant of the form in eq.
\ref{eq:dim5op}, $\Phi_D$
can be in any representation included in the symmetric product of
two adjoint representations of the group. 
When  $\Phi_{D}$ develops a 
vacuum expectation value ($vev$) $v_D$, which breaks the GUT
symmetry and sets the scale of
grand unification $M_X$, an effective gauge kinetic term is
generated from eq. \ref{eq:dim5op}. Depending on the 
structure of the $vev$, this additional contribution
usually will not be the same for the different subgroups to which
the GUT group is broken,
leading, after a scaling of the gauge
fields, to a  modification of the unification condition to:
\begin{equation}
g_{i}^{2}(M_{X})(1+\epsilon\delta_i)= g_{U}^2  ,
\label{eq:modunif}
\end{equation}
wherein $g_U$ is the unified gauge coupling, $\epsilon = \eta
v_{D}/2M_{Pl} \sim {\cal O} (M_X/M_{Pl})$, and the
group-theoretic factors $\delta_i$ arise from eq.
\ref{eq:dim5op}. The $\delta_i$ were available in the literature
for some selected choices of $\Phi_D$ and GUT groups \cite{prev}.
They were exhaustively evaluated for the first time for all
possible $\Phi_D$ for $SU(5), ~SO(10)$ and $E(6)$
GUTs\footnote{In SUSY the $\delta_i$ also have a direct
application in the non-universality of gaugino masses
\cite{cr, martin, bc}.} in \cite{cr}.

While normally in 
GUTs the gauge couplings are expected to reach a common value at
$M_X$ \cite{beta}, in the presence of dim-5 terms, as in eq.
\ref{eq:dim5op}, the modified boundary conditions of
eq. \ref{eq:modunif} must be satisfied. It is indeed possible
that this tweaking will be just enough to entail the
unification programme to succeed with the current low energy
values of the coupling constants as a boundary condition. To
check this for $SU(5)$, $SO(10)$, and $E(6)$-based GUT models is
the main goal of this work. We discuss both the
non-supersymmetric and supersymmetric  alternatives.

For $SU(5)$ this analysis has appeared in our earlier short note
\cite{cr} and it is briefly recapitulated here. GUTs based on
$SO(10)$ and $E(6)$ provide several routes of descent to the SM,
different levels of symmetry being active at the intermediate
stages. This richer structure often bears new testable features.
One of these is the possibility of $n-\bar{n}$ oscillations which
in $SO(10)$ can be mediated {\em via} scalar fields that are not
superheavy. Also, the right-handed neutrino, $\nu_R$, which is present in
both $SO(10)$ and $E(6)$ GUTs, can lead to light neutrinos
through the see-saw mechanism. If the neutrino Yukawa couplings
are not unnaturally small, the see-saw mechanism posits a large
Majorana mass for the $\nu_R$. This mass is fixed by the scale of
$(B-L)$ symmetry breaking which is determined in our analyses below.

For $SO(10)$ we examine the breaking through the intermediate
Pati-Salam (${\mathcal G}_{224} \equiv SU(2)_L \otimes SU(2)_R
\otimes SU(4)_c$) symmetry. ${\mathcal G}_{224}$ itself can break
directly to the SM or {\em via} another intermediate group ${\mathcal
G}_{2131} \equiv SU(2)_L
\otimes U(1)_{R}\otimes SU(3)_c\otimes U(1)_{(B-L)}$.  We explore
both routes.  $E(6)$ allows an intermediate $SO(10)$
symmetry and in this case the results are to a great extent
similar to that of $SO(10)$ GUTs. Here we look at $E(6)$ breaking
{\em via} the intermediate gauge group ${\mathcal G}_{333} \equiv
SU(3)_L \otimes SU(3)_R\otimes SU(3)_c$ with possibly also an
intervening ${\mathcal G}_{21213} \equiv SU(2)_L \otimes
U(1)_{Y'_{L}}\otimes SU(2)_R \otimes U(1)_{Y'_{R}} \otimes
SU(3)_c$ symmetry before descending to the SM.

When there are intermediate scales in the GUT symmetry breaking
the scalar masses have been fixed using the `{\em
Extended Survival Hypothesis}' (ESH) \cite{ESH} which is motivated along the
following lines. Normally, the lack of any protection mechanism
will tend to move all scalar masses to the GUT scale. The
necessity of light scalars is dictated by the requirement to
trigger spontaneous symmetry breaking at lower energies and this
entails a fine tuning in the scalar sector. The `Extended
Survival Hypothesis', which can also be termed `Minimal
Fine Tuning', simply requires that all scalars acquire mass at
the GUT scale barring those that are essential for symmetry
breaking at lower scales. The latter carry masses of the order of
the scales of the symmetry breakings for which they are
responsible. For any such scalar, at intermediate stages of
symmetry above its mass-scale, out of the full GUT scalar
multiplet only the submultiplet containing this scalar remains at
that scale, the remainder being at $M_X$. As an illustrative
example consider the decay chain
\begin{equation}
SO(10)\stackrel{M_X}{\longrightarrow} SU(2)_L \otimes SU(2)_R \otimes
SU(4)_c \stackrel{M_R}{\longrightarrow} SU(2)_L
\otimes U(1)_Y \otimes  SU(3)_c 
\stackrel{M_Z}{\longrightarrow} U(1)_{em} \otimes SU(3)_c.
\end{equation}
The electroweak symmetry breaking is through the ${\mathcal
G}_{SM}$ doublets, $(2,\pm 1,1)$, which emerge from a ${\mathcal
G}_{224}$ submultiplet (2,2,1) which is a part of the $SO(10)$
multiplet 10. Under ${\mathcal G}_{224}$, 10 $\equiv$ (1,1,6) +
(2,2,1). According to the ESH, out of the 10 of $SO(10)$ the
scalars forming the (1,1,6)  submultiplet  
acquire a mass $M_X$, while the (2,2,1) under ${\mathcal
G}_{224}$ are at the electroweak scale $M_Z$. The scalar
masses determine from which energy their effect on gauge coupling
evolution has to be included. Whenever
earlier work including the ESH contribution to gauge coupling
evolution is available with which our results can be compared, we
do so.
\begin{table}
\begin{center}
\begin{tabular}{|c|c|c|c|} \hline
$SU(5)$  Representations & $\delta_1$ & $\delta_2$ & $\delta_3$  \\
\hline 
{\bf 24}  & 1/$\sqrt{15}$ & 3/$\sqrt{15}$ & -2/$\sqrt{15}$ \\
{\bf 75}  &  4/$\sqrt{3}$  & -12/5$\sqrt{3}$ & -4/5$\sqrt{3}$\\
{\bf 200}  &  1/$\sqrt{21}$   &    1/5$\sqrt{21}$ &   1/10$\sqrt{21}$\\
\hline
\end {tabular}
\caption{Effective contributions ($\delta_i$) to gauge kinetic terms from
different Higgs representations in eq. \ref{eq:dim5op} for $SU(5)$.
(see eq. \ref{eq:modunif}.)}
\label{tab:su5}
\end {center}
\end{table}

The generic RG equations governing gauge coupling evolution are:
\begin{equation}
\mu \frac{dg_i}{d\mu} = \beta_i(g_i, g_j), \;\;\;\;\; (i,j = 1,\ldots,n),
\end{equation}
where $n$ is the number of couplings in the theory and at two-loop order
\begin{equation}
\beta_{i}(g_i, g_j)=(16\pi^{2})^{-1}b_{i}g_{i}^{3} 
          +(16\pi^{2})^{-2} \sum_{j=1}^{n} b_{ij}g_{j}^{2}g_{i}^{3}.
\label{eq:RG}
\end{equation}
When using this two-loop formula,
the matching of the coupling constant $\alpha_k$ below an
intermediate scale $M_I$ which goes over to $\alpha_l$
thereafter follows the relation \cite{hall2lp, oldso10}:
\begin{equation}
\frac{1}{\alpha_k(M_I)} - \frac{C_k}{12 \pi} =
\frac{1}{\alpha_l(M_I)} - \frac{C_l}{12 \pi} 
\label{eq:match} 
\end{equation}
where $C_k$ is the quadratic Casimir for the $k$-th subgroup. At
the unification scale, $M_X$, this has to be supplemented with the
contributions from the dim-5 operators  in 
eq. \ref{eq:modunif}. 


A subtle feature \cite{mix1, mix2}, considered most recently
within the context of $SO(10)$ in \cite{Berto}, has to do with
the dynamical mixing of two $U(1)$ subgroups of an intermediate
gauge symmetry even at the one-loop level. The $U(1)$ gauge
currents and the $U(1)$ gauge boson fields are by themselves
gauge invariant and so cross-couplings between them are not
forbidden by gauge symmetry. Even if the mixing is set to zero at
some scale it emerges again through the RG flow. The origin of this
mixing in the RG equations lies in the following fact:
while the trace of the product of two different $U(1)$ generators
vanishes over an entire gauge multiplet, when only a submultiplet
is light (e.g., some scalars of a multiplet remaining light due
to the Extended Survival Hypothesis in $SO(10)$ or  $E(6)$,  or
incomplete light fermion multiplets in $E(6)$) this is no longer
so. This requires a more sophisticated analysis leading to a
coupling of $g_{1m}$ and $g_{1n}$ in the one- and two-loop RG equations
where $m$ and $n$ identify two $U(1)$
groups.  These terms, not made explicit in eq. \ref{eq:RG}, arise
in the two-step breaking options for $SO(10)$ and $E(6)$ and
are detailed in the discussions in the respective sections.
 
We consider both non-supersymmetric as well as supersymmetric
versions of the theory. In the latter case the contributions of
the superpartners to the beta functions are included. (We assume
that the SUSY scale is at $M_{SUSY}$ = 1 TeV.) As is
well-known \cite{susygut}, unification of coupling constants is
compatible with  TeV-scale supersymmetry. We find that 
addition of the dim-5 contributions does not spoil this.

This paper is structured as follows. In the next section we
recapitulate the case of $SU(5)$ GUTs to set the {\em modus
operandi} for the programme. In the two subsequent sections we
consider  $SO(10)$- and $E(6)$-based theories where we also
explore the possibility of one or more intermediate mass scales.
We require that the unification scale be above the lower bound
from proton decay\footnote{The current bound \cite{pdg} $\tau_p
(p \rightarrow e^+ \pi^0) > 1.6 \times 10^{33}$ years translates
to $M_X > 10^{15.4}$ GeV. Conservatively, we use a lower limit of
$10^{16}$ GeV for $M_X$.} and below the Planck scale and that all
couplings should remain perturbative throughout the energy range.
We find that in most cases there is one intermediate scale which
can be as low as 1-10 TeV at which one expects a $Z'$ neutral
gauge boson and possibly other new particles. These provide a
testable prediction within striking range of the LHC. The other
scale(s) populating the GUT desert  are usually high and
$n-\bar{n}$ oscillations may not be observable\footnote{For one
exceptional case, see subsection \ref{s:so10_1dv}.}. In the final
section we summarise the results.

\section{$SU(5)$}\label{s:su5}
The group $SU(5)$ supports the leanest grand unified theory. It
incorporates the quarks and leptons of one generation in two irreducible
representations: $\bar{5}$ and 10.  Unlike $SO(10)$ and $E(6)$,
which are groups of rank 5 and 6 respectively, $SU(5)$ being  a
group of rank 4, it only permits a direct breaking to the SM with
no intermediate step possible. Though one of our aims in this
work is to look for intermediate scales in GUT symmetry breaking,
for the sake of completeness we give a brief account of the
results for $SU(5)$ \cite{prev,cr}. The symmetry breaking is:
\begin{equation}
SU(5) \stackrel{M_X}{\longrightarrow}  
SU(2)_L \otimes U(1)_Y \otimes SU(3)_c .
\end{equation}
The adjoint representation of $SU(5)$ is 24-dimensional. Since 
$(24 \otimes 24)_{sym} = 1 \oplus  24 \oplus 75 \oplus  200 $,
non-trivial contributions in eq. \ref{eq:dim5op} can arise if
$\Phi_D$ transforms as the 24, 75, or 200 representation. The
deviations from gauge unification due to these representations,
parametrised by the $\delta_i$ in eq. \ref{eq:modunif}, are
listed in Table \ref{tab:su5}.
\begin{table}
\begin{center}
\begin{tabular}{|c|c|c|c|c|}
\hline
{$SU(5)$}&\multicolumn{2}{|c|}{Non-SUSY}&\multicolumn{2}{|c|}{SUSY}
\\ \cline{2-5}
{representations} & $\epsilon$ & $M_X$ (GeV)& $\epsilon$ & $M_X$ (GeV)  \\
\hline 
{\bf 24}& 0.077 &4.78$\times 10^{13}$ & -0.009 &1.64$\times 10^{16}$   \\
{\bf 75}  & -0.039 &2.37  $\times 10^{15}$& 0.004 &1.22  $\times 10^{16}$ \\
{\bf 200} & -1.27 & 2.59 $\times 10^{17}$& 0.146 & 9.35 $\times 10^{15}$  \\
\hline
\end {tabular}\\
\caption{$SU(5)$ dimension-5 interaction strength, $\epsilon$, and the
gauge unification scale,  $M_X$, for different $\Phi_D$
representations using the two-loop RG equations.}
\label{tab:su5uni}
\end {center}
\end{table}
The evolution of the gauge
couplings\footnote{$g_{1,2,3}$ correspond to the $U(1)_Y, ~SU(2)_L$
and $SU(3)_c$ subgroups, respectively.} are governed by the one-
and two-loop beta-function coefficients: 
\begin{equation}
 b_{1}=4 + {1\over{10}}n_{H};\;\;\;
b_{2}=-10/3 + {1\over6}n_{H}; \;\;\;
b_{3}= -7;
\label{eq:1lsm}
\end{equation}
and
\begin{equation}
b_{ij} =
\left( \begin{array}{ccc}
19/5 & 9/5 & 44/5 \\
3/5 & 11/3 & 12 \\
11/10 & 9/2 & -26 \\
\end{array}
\right)
+n_{H}\left( \begin{array}{ccc}
9/50 & 9/10 & 0 \\
3/10 & 13/6 & 0 \\
0 & 0 & 0 \\
\end{array}
\right).
\label{eq:2lsm}
\end{equation}
$n_H$ (=1 for the SM) being the number of Higgs
doublets.
These are for the non-supersymmetric case. 

For SUSY one must also include the contributions from the
superpartners to the beta-function coefficients. With three
generations and two Higgs doublets one has:
\begin{equation}
 b_{1}={33\over 5};\;\;
b_{2}={1}; \;\;
b_{3}={-3};\;\;\;\;
b_{ij} =\left( \begin{array}{ccc}
199/25 & 27/5 & 88/5 \\
9/5 & 25 & 24 \\
11/5 & 9 & 14 \\
\end{array}
\right).
\label{eq:susy}
\end{equation}
Below $M_{SUSY}$, eqs.
\ref{eq:1lsm} and \ref{eq:2lsm} are operative with $n_H$=2
while beyond $M_{SUSY}$ eq.  \ref{eq:susy} is employed.

The results of a two-loop RG 
analysis are shown in Table \ref{tab:su5uni}. We find that for
both the non-SUSY as well as the SUSY alternatives unification is
possible in the $SU(5)$ GUT when additional effective
interactions of dimension-5 are in play.  $M_X$, the unification
scale, and $\epsilon$, the strength of the
dim-5 interaction, are shown in Table
\ref{tab:su5uni} for the different choices of $\Phi_D$. It is seen
that for the non-SUSY case, unification, though achievable with
the dim-5 interactions, is not satisfactory. For $\Phi_{24}$ and
$\Phi_{75}$ the unification scale $M_X$ is too low to be
consistent with the current limits on the proton decay lifetime
while for $\Phi_{200}$ $\epsilon$ is larger than unity. The
solutions for the SUSY case are satisfactory on every
count.

\section{$SO(10)$}
$SO(10)$ \cite{so10} is the smallest GUT which accommodates all
the fermions of a generation in one irreducible multiplet, the
spinorial 16.  The group admits a left-right symmetric subgroup
\cite{lrs} -- the Pati-Salam $SU(2)_L \otimes SU(2)_R \otimes
SU(4)_c $ which we denote by ${\mathcal G}_{224}$ -- with
interesting new phenomenology including quark-lepton unification
within the $SU(4)_c$.
The chain of symmetry breaking that we discuss here is
\begin{equation}
SO(10)\stackrel{M_X}{\longrightarrow} SU(2)_L \otimes SU(2)_R \otimes SU(4)_c 
\stackrel{M_C}{\longrightarrow}
SU(2)_L \otimes U(1)_{R} \otimes SU(3)_c \otimes U(1)_{(B-L)} 
\stackrel{M_R}{\longrightarrow} SM.
\label{so10chain}
\end{equation}
Some subcases which we also look at are when (i) $M_X = M_C =
M_R$ which corresponds to a breaking of $SO(10)$ to the SM with
no intervening steps, and (ii) $M_C = M_R$ which is a situation
where $SO(10)$ reduces to the SM through one intermediate step.
We consider these cases one by one. All results presented below
are based on two-loop RG analyses.  

The adjoint representation of $SO(10)$ is 45-dimensional. Since
$(45 \otimes 45)_{sym} = 1 \oplus  54 \oplus 210 \oplus  770 $,
$\Phi_D$ in eq. \ref{eq:dim5op} transforms as the 54, 210, or 770
representation. The deviations from gauge unification due to
these representations, parametrised by the $\delta_i$ in eq.
\ref{eq:modunif}, are listed in Table \ref{tab:so10}.
\begin{table}
\begin{center}
\begin{tabular}{|c|c|c|c|} \hline
$SO(10)$  Representations  & $\delta_{2L}$ & $\delta_{2R}$& $\delta_{4c}$  \\
\hline 
{\bf 54}   & 3/$2\sqrt{15}$ & 3/$2\sqrt{15}$& -1/$\sqrt{15}$\\
{\bf 210}   & 1/$\sqrt2$ & -1/$\sqrt2$&  0\\
{\bf 770}   &  5/$3\sqrt5$   & 5/$3\sqrt5$&   2/$3\sqrt5$\\
\hline
\end {tabular}
\caption{Effective contributions ($\delta_i$)  to gauge kinetic terms from
different Higgs representations in eq. \ref{eq:dim5op} for
$SO(10)$ \cite{cr}.  (see eq. \ref{eq:modunif}.)}
\label{tab:so10}
\end {center}
\end{table}
\subsection{No-step breaking  in $SO(10)$} 
This is the most straight-forward symmetry breaking for $SO(10)$
and is much like the $SU(5)$ case discussed in section \ref{s:su5}.
\begin{equation}
SO(10) \stackrel{M_X}{\longrightarrow}SU(2)_L
\otimes U(1)_Y \otimes  SU(3)_c .
\end{equation}
When there are no intermediate scales the gauge coupling
evolutions are governed by eqs. \ref{eq:1lsm} and
\ref{eq:2lsm} for the non-supersymmetric case and eq. 
\ref{eq:susy} for the SUSY version.   

\begin{table}[hbt]
\begin{center}
\begin{tabular}{|c|c|c|c|c|}
\hline
$SO(10)$&\multicolumn{2}{|c|}{Non-SUSY}&\multicolumn{2}{|c|}{SUSY} \\ \cline{2-5}
representations & $\epsilon$ & $M_X$ (GeV)& $\epsilon$ & $M_X$ (GeV)  \\
\hline 
{\bf 54} & 0.170 &3.99$\times 10^{13}$ & -0.013 &1.54$\times 10^{16}$   \\
{\bf 210}  & 0.088 &4.39  $\times 10^{14}$& -0.008 &1.35  $\times 10^{16}$ \\
{\bf 770}  & 0.274 & 4.10 $\times 10^{13}$& -0.018 & 1.54 $\times 10^{16}$  \\
\hline
\end {tabular}\\
\caption{Dimension-5 interaction strength, $\epsilon$, and the
gauge unification scale,  $M_X$, for different $\Phi_D$
representations using two-loop RG equations when $SO(10)$
descends directly to the SM.}
\label{t:0iso10}
\end {center}
\end{table}

The results are shown in Table \ref{t:0iso10}. As for $SU(5)$, we
find that the non-supersymmetric solutions are untenable. For all
three choices of $\Phi_D$ the unification scale is
$\mathcal{O}(10^{13} - 10^{14})$ GeV, which is excluded by the
current observational bounds on the proton decay lifetime.

\subsection{One-step breaking in $SO(10)$} 
\noindent
Here we have to consider the following breaking chain of $SO(10)$
\begin{equation}
SO(10)\stackrel{M_X}{\longrightarrow} SU(2)_L \otimes SU(2)_R \otimes
SU(4)_c \stackrel{M_C}{\longrightarrow} SM.
\label{so1stp}
\end{equation}
The ${\mathcal G}_{224}$ intermediate group offers a new
discrete symmetry -- D-parity \cite{dpar, oldso10}. This symmetry
relates the gauged $SU(2)_L$ and $SU(2)_R$ subgroups of $SO(10)$
much the same way that ordinary Parity relates the $SU(2)_L$ and
$SU(2)_R$ subgroups of the Lorentz group $SO(3,1)$. Alternative
routes of $SO(10)$ symmetry breaking are admissible which either
preserve or violate D-parity at the intermediate stages. We will
consider both in the following.  The first step of
symmetry breaking from $SO(10)$ to ${\mathcal G}_{224}$ is
accomplished by assigning an appropriate $vev$ to a 54, 210, or
770-dimensional Higgs.  $\langle \Phi_{54} \rangle$ or $\langle
\Phi_{770} \rangle$ ensure that D-parity is conserved while $\langle
\Phi_{210} \rangle$ breaks D-parity.  This is reflected in Table
\ref{tab:so10} in that $\delta_{2L} = -\delta_{2R}$ in this case
whereas in the other cases they are equal. 

The next step breaking of ${\mathcal G}_{224}$ to the SM is
achieved through the $vev$ of a 126-dimensional Higgs. The
submultiplet of $126_H$ that develops a $vev$ for this purpose at
the scale $M_C$ transforms as (1,3,$\overline{10}$) under
${\mathcal G}_{224}$. According to the Extended Survival
Hypothesis the entire submultiplet acquires a mass
$\mathcal{O}(M_C)$ while the other members of $126_H$ are at
$M_X$. This is true if D-parity is not conserved. When D-parity
remains unbroken then it relates the (1,3,$\overline{10}$)
submultiplet to the ($3$,1,10) $ \subset 126_H$ and it too has a
mass of $\mathcal{O}(M_C)$.

One must also consider the Higgs scalars $\phi_{SM}$ responsible
for the breaking of SM at $\sim M_Z$. They transform under
${\mathcal G}_{SM},  ~{\mathcal G}_{224}$, and $SO(10)$ as \{(2,1,1)
+ (2,-1,1)\}, (2,2,1) and 10, respectively. Notice that the
Extended Survival Hypothesis mandates that the (1,1,6) under
${\mathcal G}_{224}$ contained in the $SO(10)$ 10-dimensional
representation has a mass at $M_X$ while the (2,2,1) is at $M_Z$.

The scalars contributing to the RG evolution in different stages
are summarised in Table \ref{t:esh10s1}.
\begin{table}[hbt]
\begin{center}
\begin{tabular}{|c|c|c|c|}
\hline
$SO(10)$&Symmetry&\multicolumn{2}{|c|}{Scalars contributing to RG}\\ \cline{3-4}
representation & breaking & $M_Z \rightarrow M_C$ & $M_C \rightarrow M_X$ \\
 & & Under ${\mathcal G}_{SM}$ & Under ${\mathcal
G}_{224}$   \\ \hline 
{\bf 10} & ${\mathcal G}_{SM} \rightarrow EM$ & (2,$\pm1$,1) &
(2,2,1)\\  
& & & \\
{\bf 126} & ${\mathcal G}_{224} \rightarrow {\mathcal G}_{SM}$ &
-  & (1,3,$\overline{10}$)\\  
 & &
  & \{(3,1,10)\} \\  \hline
\end {tabular}
\caption{Higgs scalars for the one-step symmetry breaking of
$SO(10)$ and the submultiplets contributing to RG evolution
according to the ESH. The submultiplet in the braces also
contributes if D-parity is conserved.  }
\label{t:esh10s1}
\end {center}
\end{table}

When the couplings are evolved from their low energy inputs
the key matching formula at $M_C$ is\footnote{$\alpha_{1Y}$ is
the GUT-normalised $U(1)_Y$ coupling.}:
\begin{equation}
\frac{1}{\alpha_{1Y}(M_C)} = \frac{3}{5}\left[
\frac{1}{\alpha_{2R}(M_C)} - \frac{1}{6\pi} \right]+ 
\frac{2}{5}\left[\frac{1}{\alpha_{4c}(M_C)}  - \frac{1}{3\pi}\right] \;\; .
\label{e:match10i1}
\end{equation} 
This is a consequence of the relation $Y/2 = T_{3R} + (B-L)/2$.
On the r.h.s. $T_3$ resides within the $SU(2)_R$ while $(B-L)$ is
included in $SU(4)_c$ and eq. \ref{eq:match} has been used.
Similarly, 
${\alpha_{4c}(M_C)} = {\alpha_{3c}(M_C)} + 1/12\pi$ and is fixed from the
RG evolution of ${\alpha_{3c}}$ from  $M_Z$. The two
cases that we discuss here are:

(a) If D-parity is not conserved then  for every choice of $M_C$,
eq. \ref{e:match10i1} determines $\alpha_{2R}(M_C)$. The three
couplings have to be further evolved to determine $M_X$ and $\epsilon$.

(b) If D-parity is conserved at $M_C$ then in eq.
\ref{e:match10i1} we must further impose $\alpha_{2R}(M_C) =
\alpha_{2L}(M_C)$, with the latter fixed by the RG evolution of
${\alpha_{2L}}$ from its low energy value. This identifies a
unique $M_C$. $M_X$ can then be determined in terms of $\epsilon$.

We discuss these options in detail below.

{\bf From $M_Z$ to $M_C$}: For the RG running of the coupling
constants in this range eqs. \ref{eq:1lsm}, \ref{eq:2lsm}, and
\ref{eq:susy} are applicable irrespective of whether D-parity is
conserved or not.

\subsubsection{D-parity not conserved}\label{s:so10_1dv}
\noindent
This is the case when $\Phi_{210}$ is responsible for the
$SO(10)$ GUT symmetry breaking.

{\bf From $M_C$ to $M_X$}:

The beta-function coefficients receive
contributions from (1,3,$\overline{10}) \subset 126_H$ along with
the (2,2,1)$\subset 10_H$ scalars and the three generations of
fermions: $(2,1,4)+(1,2,\bar{4}) = 16_F$. These are:
\begin{equation}
{\rm \mbox{NON-SUSY:}}~~ b_{2L}=-3; \;\;
b_{2R}=11/3; \;\;
b_{4c}=-23/3;\;\;\;\;b_{ij} =\left( \begin{array}{ccc}
8 & 3 & 45/2 \\
3 & 584/3 & 765/2 \\
9/2 & 153/2 & 643/6 \\
\end{array}
\right).
\label{eq:oddso10UR}
\end{equation}

\begin{equation}
{\rm \mbox{SUSY:}}~~  b_{2L}=1; \;\;
b_{2R}=21; \;\;
b_{4c}=3;\;\;\;\;
b_{ij} =\left( \begin{array}{ccc}
25 & 3 & 45 \\
3 & 265 & 405 \\
9 & 81 & 231 \\
\end{array}
\right).
\label{eq:oddso10URS}
\end{equation}
The one- and two-loop beta-function coefficients we have
calculated are in agreement\footnote{There are minor differences
in $b_{2L2R}$ and $b_{2L4c}$ between our results and that in \cite{oldso10}.}
with those obtained in \cite{oldso10, pbp1}. Both papers
deal only with the non-SUSY case.

\begin{center}
\begin{figure}[thb]
\hskip .40cm
\includegraphics[width=5.1cm,height=4.cm,angle=0]{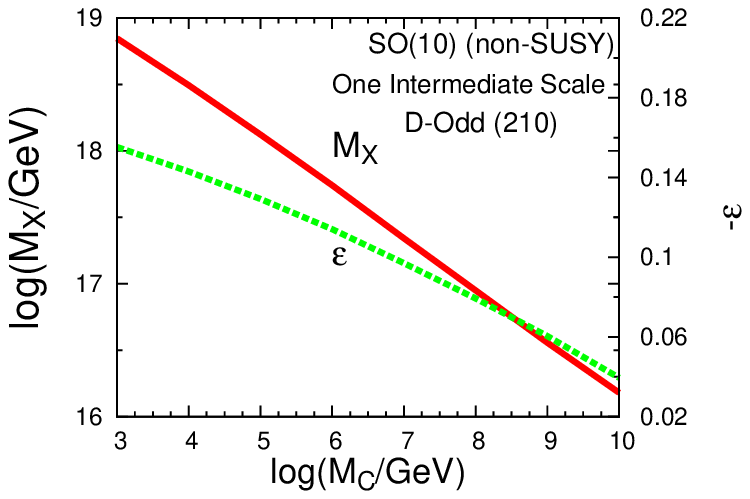}
\hskip 0.1cm
\includegraphics[width=5.9cm,height=4.0cm,angle=0]{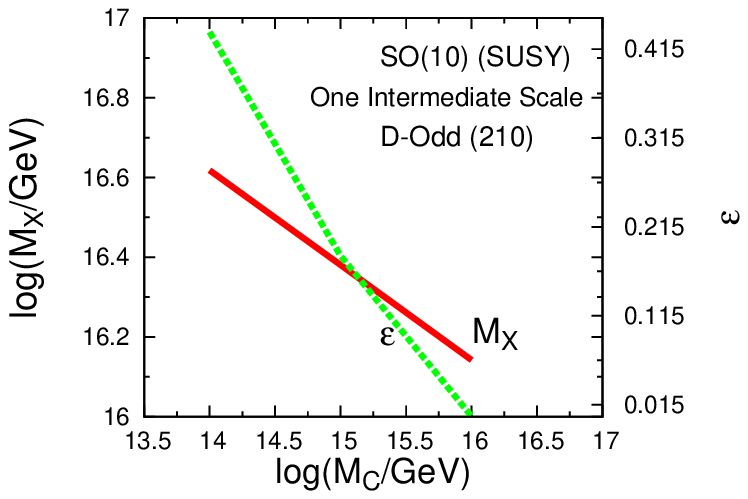}
\hskip 0.1cm
\includegraphics[width=5.3cm,height=4.0cm,angle=0]{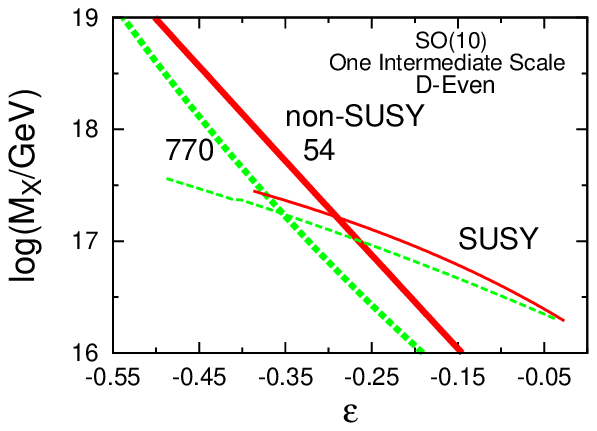}
\caption{\sf \small {$SO(10)$ one-step breaking results: 
The unification scale, $M_X$, (red dark solid lines) and
the strength of the dim-5 interaction, $\epsilon$, (green pale broken
lines) as a function of $M_C$ for the D-parity
nonconserving  ($\Phi_{210}$) case for (left) non-SUSY   and  (centre) SUSY.
$M_X$ vs. $\epsilon$ for the D-parity
conserving case (right).  Thick (thin) lines correspond to non-SUSY
(SUSY). The results for both $\Phi_{54}$
(red dark solid) and $\Phi_{770}$ (green pale broken) are shown.}}
\label{f:s1ns}
\end{figure}
\end{center}

{\bf Results:} For this chain, the low energy
measured gauge couplings allow a range of values for $M_C$. The
results for this case are shown in the left (non-SUSY) and middle
(SUSY) panels of Fig. \ref{f:s1ns}. As shown, for
every allowed $M_C$ one can determine $M_X$ (red dark solid curve) and
$\epsilon$ (green pale broken curve) from the unification of coupling
constants satisfying eq. \ref{eq:modunif}.  As a general
observation, lower values of $M_C$ correspond to increased $M_X$
and larger $\epsilon$. Notice that in the non-SUSY case, $M_C$
can be as low as $10^3$ GeV and therefore within the range of
detectability for the Large Hadron Collider. Further, the
(1,3,$\overline{10}$) scalars which have mass $\sim M_C$ can
mediate $n-\bar{n}$ oscillations\footnote{The oscillation period
$\tau_{n-\bar{n}} \sim (M_{(1,3,\overline{10})})^5$.}  and it is
known that current experimental limits place a lower bound on
$M_C$ around 10 TeV depending on hadronic factors not precisely
known \cite{nnbar}. The mass of the $\nu_R$ is also ${\cal
O}(M_C)$. While a low $M_C$ is desirable for
detectability of $n-\bar{n}$ oscillations it is not the preferred
choice for a see-saw mechanism for generating light neutrino
masses.  In the SUSY case $M_X$ and $M_C$ are
restricted to a very limited range, a reflection of the large
beta functions beyond $M_C$.  Here $M_C$ ($10^{14} - 10^{16}$
GeV) is too high for observable $n-\bar{n}$ oscillations but
quite appropriate for light neutrino see-saw masses.

\subsubsection{D-parity conserved}
\noindent

This is the situation which arises when either $\Phi_{54}$ or
$\Phi_{770}$ is responsible for the $SO(10)$ breaking.

{\bf From $M_C$ to $M_X$}:

According to the Extended Survival
Hypothesis the only change from the previous subsection is that
one must include
contributions from both (1,3,$\overline{10}$) and (3,1,10)
within the $126_H$.  
This gives:
\begin{equation}
{\rm \mbox{NON-SUSY:}}~~ b_{2L}=b_{2R}=11/3; \;\;\
b_{4c}=-14/3;\;\;\;\;\;b_{ij} =\left( \begin{array}{ccc}
584/3 & 3 & 765/2 \\
3 & 584/3 & 765/2 \\
153/2 & 153/2 & 1759/6 \\
\end{array}
\right).
\label{eq:evenso10}
\end{equation}

\begin{equation}
{\rm \mbox{SUSY:}}~~ b_{2L}=b_{2R}=21; \;\;
b_{4c}=12;\;\;\;\;b_{ij} =\left( \begin{array}{ccc}
265 & 3 & 405 \\
3 & 265 & 405 \\
81 & 81 & 465 \\
\end{array}
\right).
\label{eq:evenso10S}
\end{equation}

The beta-function coefficients for the non-SUSY case agree with 
those in \cite{pbp1}. 

{\bf Results:} In this case, the relationship between the $SU(2)_L$ and
$SU(2)_R$ couplings uniquely fix the intermediate scale $M_C$. We
find that for the non-SUSY case $M_C =$ 5.37 $\times
10^{13}$ GeV while in the SUSY case it is higher and is around
1.9 $\times 10^{16}$ GeV. This fixed intermediate scale, $M_C$,
is the same for $\Phi_{54}$ and $\Phi_{770}$. The
(1,3,$\overline{10}$) and (3,1,10) scalars at $\sim M_C$ are thus
too heavy for observable
$n-\bar{n}$ oscillations.
Depending on
whether the non-SUSY or the SUSY theory is under consideration, a
range of allowed  $M_X$ can be obtained as a function of
$\epsilon$ for either choice of $\Phi_D$.
The results for the non-SUSY (thick lines) and SUSY (thin
lines) cases are shown in the right panel of Fig.  \ref{f:s1ns}.
The dark solid (red) lines correspond to $\Phi_{54}$ while the
pale broken (green) lines are for $\Phi_{770}$.

\subsection{Two-step breaking in $SO(10)$}
\noindent

Here we consider the breaking of $SO(10)$ to SM {\em via} two
intermediate steps:
\begin{equation}
SO(10)\stackrel{M_X}{\longrightarrow} SU(2)_L \otimes SU(2)_R \otimes
SU(4)_c \stackrel{M_C}{\longrightarrow} SU(2)_L \otimes U(1)_{R}
\otimes SU(3)_c \otimes U(1)_{(B-L)} \stackrel{M_R}{\longrightarrow}
SM.
\label{so102s}
\end{equation}
The symmetry breaking at different stages is arranged as follows.
The breaking of the Pati-Salam ${\mathcal G}_{224}$ to ${\mathcal
G}_{2131}$ is through the $vev$ of a (1,3,15) component of
$210_H$. The subsequent descent to the  SM is through the $vev$
to a (1,3,1,-2) $\subset$ (1,3,$\overline{10}) \subset 126_H$.
 The Higgs scalars responsible for the SM symmetry breaking,
$\phi_{SM}$, transform as (2,$\pm$1,1) under the SM group and as
(2,$\pm\frac{1}{2},1,0) \subset (2,2,1) \subset$ 10 under
${\mathcal G}_{2131}$, ${\mathcal G}_{224}$, and $SO(10)$,
respectively.  The contributing scalars at different stages
of RG evolution, as determined by the ESH, are summarised in Table
\ref{t:esh10s2}.

\begin{table}[hbt]
\begin{center}
\begin{tabular}{|c|c|c|c|c|}
\hline
$SO(10)$&Symmetry&\multicolumn{3}{|c|}{Scalars contributing to RG}\\ \cline{3-5}
representation & breaking & $M_Z \rightarrow M_R$ & $M_R \rightarrow M_C$
& $M_C \rightarrow M_X$ \\
 & & Under ${\mathcal G}_{SM}$ & Under ${\mathcal
G}_{2131}$ & Under ${\mathcal
G}_{224}$   \\ \hline
{\bf 10} & ${\mathcal G}_{SM} \rightarrow EM$ & (2,$\pm1$,1) & 
(2,$\pm\frac{1}{2}$,1,0) &
(2,2,1)\\  
& & & & \\
{\bf 126} & ${\mathcal G}_{2131} \rightarrow {\mathcal G}_{SM}$ &
-  & (1,3,1,-2) & (1,3,$\overline{10}$)\\  
 & &
  & & \{(3,1,10)\} \\  
& & & & \\
{\bf 210} & ${\mathcal G}_{224} \rightarrow {\mathcal G}_{2131}$ &
-  & - & (1,3,15)\\  
 & &
  & & \{(3,1,15)\} \\  \hline
\end {tabular}\\
\caption{Higgs scalars for the two-step symmetry breaking of
$SO(10)$ and the submultiplets contributing to RG evolution
according to the ESH. The submultiplets in the braces also
contribute if D-parity is conserved.  }
\label{t:esh10s2}
\end {center}
\end{table}

If D-parity is conserved, and it can be conserved only till
$M_C$ in this chain, then one must include the
contribution from a (3,1,10) and a (3,1,15) in the
final stage of evolution (see Table \ref{t:esh10s2}).

A point worth noting in Table \ref{t:esh10s2} is that in the
range $M_C$ to $M_X$ there are contributions from (1,3,15) (and
possibly (3,1,15)) scalar fields over and above those in the
one-step breaking case (see Table \ref{t:esh10s1}). Because of
these large-dimensional multiplets the RG evolutions are quite
different and the na\"{i}ve expectation of the two-step results
going over to the one-step one in the limit $M_R = M_C$ is
invalid.

In the energy range $M_R$ to $M_C$ there are two $U(1)$ gauge
groups. As observed in \cite{mix1, mix2} and stressed most
recently in \cite{Berto}, due to incomplete scalar multiplets
remaining light according to the Extended Survival Hypothesis  there is
a dynamical mixing between these two $U(1)$ subgroups which is
manifested in the RG evolution equations. In particular, below
the $M_R$ threshold there is one $U(1)$ coupling corresponding to
hypercharge, $Y$, while above one must consider the possibility
of a $2 \times 2$ matrix of $U(1)$ couplings, $G$:
\begin{equation}
G =\left( \begin{array}{cc}
g_{RR} & g_{RX} \\
g_{XR} & g_{XX} \\
\end{array}
\right),
\label{eq:G} 
\end{equation}
where $X \equiv (B-L)$. This is the most general form permitted
for the coupling of the gauge currents to gauge bosons which for
the $U(1)$ groups are both by themselves gauge invariant. Here,
$g_{ij}$ is the strength of the coupling of the $i$th current to
the $j$th gauge boson.  In the range $M_R$ to $M_C$ the evolution
of all elements of $G$ will occur\footnote{Due of the mixing
of the two $U(1)$ groups, the RG equations will be somewhat more
involved and are not presented. They can be found in \cite{mix2,
Berto}.}. The RG equations for $g_{RX}$ and $g_{XR}$
at the one-loop level involve one additional beta-function
coefficient, $\tilde b_{XR} = \tilde b_{RX} \propto \sum_i Q_R^i
Q_X^i$. At the two-loop level, besides the usual ones, one
requires the following independent coefficients:
\begin{enumerate}
\item $\tilde b_{RX,RR}, ~\tilde b_{XR,XX}$
\item $\tilde b_{RX,p}\;, ~\tilde b_{XR,p}$
\item $\tilde b_{p,RX}\;$.
\end{enumerate}
The first beta-coefficient in 1 appears in, among others,  the
evolution equation of $g^{}_{RX}$ as the coefficient of
$g^4_{RR}g^{}_{XX}$ while the second is readily obtainable from
the above through $R
\leftrightarrow X$. For 2 and 3 above, $p$ represents a
non-abelian subgroup of the gauge symmetry. The 
coefficient of $g^{3}_{RX}g^2_p$ ($g^{3}_{XR}g^2_p$) in the RG
equation of $g^{}_{RX}$ ($g^{}_{XR}$) is listed under 2 above.
Similarly, in 3, $\tilde b_{p,RX}$ is the coefficient of
$g^3_p(g^{}_{RR}g^{}_{XR} + g^{}_{XX}g^{}_{RX})$. For the
$SO(10)$ model we are considering, the entries in 2 and 3 turn out
to be zero. 

At the boundary $M_R$ there is freedom to choose $G$ to be upper
triangular. On RG evolution all elements will, however, become
non-zero. The matching of the elements of $G$ with the coupling
below $M_R$ and those above $M_C$ is made through projection
operators which relate the basis of evolution with the $U(1)$
gauge basis defining the groups at the boundary.

Taking all this into account, the gauge couplings evolve as follows:

\noindent

{\bf i-a) From $M_C$ to $M_X$ (D-parity not conserved)}:
\begin{equation}
{\rm \mbox{NON-SUSY:}}~~ b_{2L}=-3; \;\;
b_{2R}=41/3; \;\;
b_{4c}=-11/3;\;\;\;\;
b_{ij} =\left( \begin{array}{ccc}
8 & 3 & 45/2 \\
3 & 1424/3 & 1725/2 \\
9/2 & 345/2 & 1987/6 \\
\end{array}
\right).
\label{eq:oddso10UC}
\end{equation}

\begin{equation}
{\rm \mbox{SUSY:}}~~ b_{2L}=1; \;\;
b_{2R}=51; \;\;
b_{4c}=15;\;\;\;\;
b_{ij} =\left( \begin{array}{ccc}
25 & 3 & 45 \\
3 & 625 & 885 \\
9 & 177 & 519 \\
\end{array}
\right).
\label{eq:oddso10UCS}
\end{equation}

{\bf i-b) From $M_C$ to $M_X$ (D-parity conserved)}:
\begin{equation}
{\rm \mbox{NON-SUSY:}}~~ b_{2L}=
b_{2R}=41/3; \;\;
b_{4c}=10/3;\;\;\;\;
b_{ij} =\left( \begin{array}{ccc}
1424/3 & 3 & 1725/2 \\
3 & 1424/3 & 1725/2 \\
345/2 & 345/2 & 4447/6 \\
\end{array}
\right).
\label{eq:evenso10UC}
\end{equation}

\begin{equation}
{\rm \mbox{SUSY:}}~~ b_{2L}= 
b_{2R}=51; \;\;
b_{4c}=36;\;\;\;\;
b_{ij} =\left( \begin{array}{ccc}
625 & 3 & 885 \\
3 & 625 & 885 \\
177 & 177 & 1041 \\
\end{array}
\right).
\label{eq:evenso10UCS}
\end{equation}
{\bf ii) From $M_R$ to $M_C$}:

Below $M_C$, where the gauge group is $SU(2)_L \otimes U(1)_{R}
\otimes SU(3)_c \otimes U(1)_{(B-L)}$, there is no $L
\leftrightarrow R$ symmetry and hence there can be no D-parity.
Thus for the two cases just discussed the evolution will be
identical.  Here we are giving the decompositions of the
contributing fields under the gauge symmetry at this level:
\begin{eqnarray}
16_F &=&[2,0,3,-1/3]+[2,0,1,1]+[1,1/2,\bar{3},1/3] + \\ \nonumber
      & &+[1,1/2,1,-1]+[1,-1/2,\bar{3},1/3]+[1,-1/2,1,-1],\\ \nonumber
10_H &\supset& [2,1/2,1,0]+[2,-1/2,1,0], \;\;
126_H \supset [1,-1,1,2] \;\;.\nonumber
\end{eqnarray}
whence\footnote{The coefficients superscribed with a {\em tilde}
arise due to $U(1)$ mixing.} ($X \equiv (B-L)$)
\begin{equation}
{\rm \mbox{NON-SUSY:}}~~ b_{2L}=-3; \;
b_{RR}=14/3; \;
b_{3c}=-7; \;
b_{XX}=9/2; \;
\tilde b_{RX}= \tilde b_{XR} = -1/\sqrt{6},
\label{eq:1l-evenSO10CR}
\end{equation}
\begin{eqnarray}
b_{ij} &=& \left( \begin{array}{cccc}
8 & 1 & 12 & 3/2 \\
3 & 8 & 12 & 15/2 \\
9/2 & 3/2 & -26 & 1/2 \\
9/2 & 15/2 & 4 & 25/2 \\
\end{array}
\right); \nonumber \\ 
\tilde b_{XR,RR} &=& -2\sqrt{6};\;\;
\tilde b_{RX,XX} = -3\sqrt{6}, \;\;\;
 \tilde b_{RX,p} = \tilde b_{XR,p}
= \tilde b_{p,RX} = 0.
\label{eq:2l-evenSO10CR}
\end{eqnarray}
\begin{equation}
{\rm \mbox{SUSY:}}~~ b_{2L}=1; \;
b_{RR}=8; \;
b_{3c}=-3; \;
b_{XX}=15/2; \;
\tilde b_{RX}=\tilde b_{XR} = -\sqrt{6}/2,
\label{eq:1l-evenSSO10CR}
\end{equation}
\begin{eqnarray}
b_{ij} &=& \left( \begin{array}{cccc}
25 & 1 & 24 & 3 \\
3 & 11 & 24 & 9 \\
9 & 3 & 14 & 1 \\
9 & 9 & 8 & 16 \\
\end{array}
\right); \nonumber \\ 
\tilde b_{XR,RR} &=& -2\sqrt{6};\;\;
\tilde b_{RX,XX} = -3\sqrt{6}, \;\;\;
 \tilde b_{RX,p} = \tilde b_{XR,p}
= \tilde b_{p,RX} = 0.
\label{eq:2l-evenSSO10CR}
\end{eqnarray}

{\bf iii) From $M_Z$ to $M_R$}:

In this range eqs. \ref{eq:1lsm}, \ref{eq:2lsm}, and
\ref{eq:susy} are applicable. 

The one- and two-loop beta-function coefficients in the D-parity
conserving case agree with those obtained in \cite{pbp1} and
\cite{Berto} with the proviso that in \cite{pbp1} only one
Higgs doublet is assumed to contribute in the range $M_Z$ to
$M_R$. In addition, the $U(1)$ mixing contribution at the
one-loop level has been included only in \cite{Berto}. 

\begin{center}
\begin{figure}[thb]
\hskip 2.30cm
\includegraphics[width=5.4cm,height=4.60cm,angle=0]{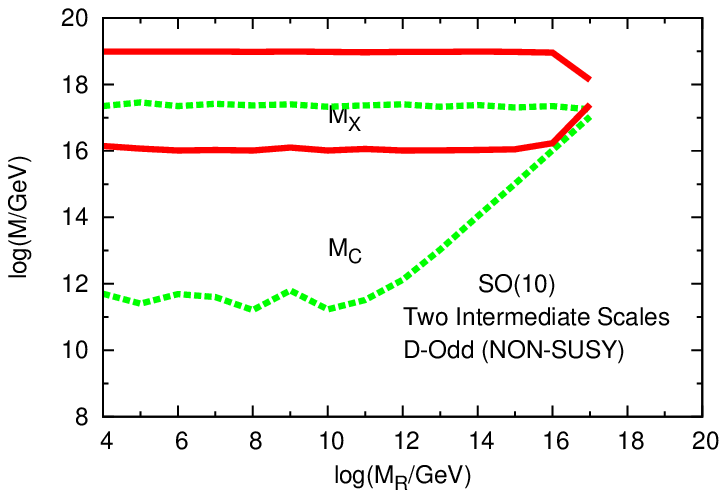}
\hskip 0.1cm
\includegraphics[width=5cm,height=4.60cm,angle=0]{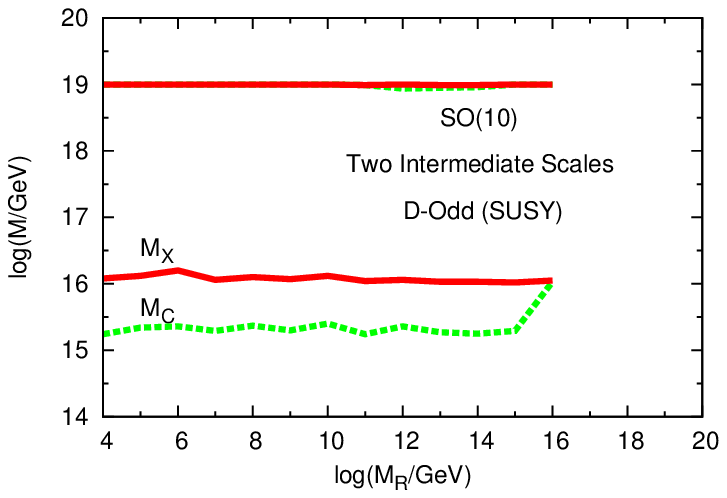}
\caption{\sf \small The allowed ranges of $M_X$ (red dark
solid) and $M_C$ (green pale broken) vs. $M_R$ for the non-SUSY
(left) and SUSY (right) cases for $SO(10)$ breaking through two
intermediate steps when D-parity is not conserved.  Note that the
upper limits for $M_X$ and $M_C$ are almost identical for SUSY.}
\label{f:sv2sn}
\end{figure}
\end{center}

{\bf Results:} 

At $M_R$ one must now use the matching relation:
\begin{equation}
\frac{1}{\alpha_{1Y}(M_R)} = 4\pi \; P \; (G \; G^T)^{-1}  P^T \;.
\label{e:match10i2}
\end{equation} 
where $P = (\sqrt{\frac{3}{5}} \;\; \sqrt{\frac{2}{5}}) $. At the
$M_C$ boundary, the $U(1)_R$ and $U(1)_{(B-L)}$ couplings are
obtained from the RG evolved $G$ using a similar formula while
choosing $P$ = (1 ~0) and (0 ~1), respectively.

When D-parity is not conserved, i.e., the first stage of symmetry
breaking is due to $\Phi_{210}$, eq. \ref{e:match10i2} fixes the
couplings at $M_R$.  The meeting of the $U(1)_{(B-L)}$ and
$SU(3)_c$ couplings determines $M_C$ and at that scale
$\alpha_{1R}$ goes over to $\alpha_{2R}$. At $M_R$, the ratios
$g_{RR}/g_{(B-L)(B-L)}$ and $g_{R(B-L)}/g_{(B-L)(B-L)}$ can be
varied to first determine $M_C$ {\em via} eq. \ref{e:match10i2}
and subsequently $M_X$. In Fig. \ref{f:sv2sn} are shown the
ranges of $M_C$ and $M_X$, consistent with all constraints,  as a
function of $M_R$ for the non-SUSY (left) and SUSY
(right) cases. Notice that in both cases $M_R$ can be as low as
10 TeV. This is the scale for a new neutral $Z'$ boson which could
be seen at the LHC. On the other hand, in both cases, $M_R$,
which is also  the $(B-L)$-violating scale relevant for see-saw
neutrino masses, can be 10$^{14-16}$ GeV, which is of the
desirable size for ${\cal O}$(1) Yukawa couplings. For the
non-SUSY case, $M_C$ is around $10^{11.5}$ GeV or above which is
too high for the detectability of $n-\bar{n}$ oscillations. For
SUSY $M_C$ is even higher, $\sim 10^{15}$ GeV or more. This is
also the mass scale for the right-handed charged gauge bosons.
For the solutions discussed above the parameter $|\epsilon |$
lies in the range (0.004-0.160) for non-SUSY and (0.04 - 1.0) for
SUSY.

When D-parity is conserved, i.e., the GUT symmetry
breaking is due to $\Phi_{54}$ or $\Phi_{770}$, $M_R$  must
be such that the  $\alpha_{1R}$ and $\alpha_{1(B-L)}$ 
matches with $\alpha_{2L}$ and 
$\alpha_{3c}$, respectively (as per eq. \ref{eq:match}) 
at precisely the same energy scale $M_C$. This is quite
constraining. Though for both non-SUSY and SUSY $M_R$ can range
from $10^4$ - $10^{16}$ GeV, $M_C$ and $M_X$ are very close to
each other\footnote{This is a consequence of the large
beta-functions due to the contributions from big submultiplets
introduced to maintain D-parity symmetry (see Table
\ref{t:esh10s2}).} and around $10^{16}$ GeV always. Thus, barring
the $Z'$ neutral gauge boson there will be no other observable
signatures in this scenario. The high values of $M_C$ preclude
the possibility of detectable $n-\bar{n}$ oscillations. On the
other hand, such a high $M_{\nu_R}$ will be able to accommodate
the light neutrino masses through a Type I see-saw. For the strength of
the dim-5 interaction, $\epsilon$, it is found $0 \leq |\epsilon
| \leq 0.18$ for non-SUSY and $0 \leq |\epsilon
| \leq 0.25$ for SUSY. For both
non-SUSY as well as SUSY, the results for
$\Phi_{54}$ and $\Phi_{770}$ are practically identical
excepting for small differences in the values of $\epsilon$.

\section{$E(6)$}
The exceptional group $E(6)$ has also been discussed in the
literature as a possible GUT symmetry \cite{e6}. 
The breaking scheme of $E(6)$ that we consider here is
\begin{equation}
E(6)\stackrel{M_X}{\longrightarrow} SU(3)_L \otimes SU(3)_R \otimes SU(3)_c 
\stackrel{M_I}{\longrightarrow}
SU(2)_L \otimes U(1)_{Y'_L} \otimes SU(2)_R \otimes U(1)_{Y'_R} \otimes SU(3)_c
\stackrel{M_R}{\longrightarrow} SM.
\label{e6chain}
\end{equation}

In $E(6)$ the fermions of one generation are accommodated in the
27-dimensional fundamental representation which under ${\mathcal
G}_{333}$ consists of ($\bar{3}$,3,1) + (1,$\bar{3}$,3) +
(3,1,$\bar{3}$). At the stage where the ${\mathcal G}_{333}$
symmetry is broken, all fermions other than those in the SM become
massive.

In contrast to the $SO(10)$ cases discussed in the previous
section, here the quark-lepton symmetry is lost at $M_X$ 
and $n-\bar{n}$ oscillations will be highly
suppressed in this class of $E(6)$ models.

The adjoint representation of $E(6)$ is 78-dimensional. Since
$(78 \otimes 78)_{sym} = 1 \oplus  650 \oplus 2430$, non-trivial
contributions in eq. \ref{eq:dim5op} can arise if $\Phi_D$
transforms as the 650 or 2430 representation. Of these, $\Phi_{650}$ has
two distinct directions for the $vev$ which can accomplish the
symmetry breaking to ${\mathcal G}_{333}$, one of
which protects D-parity while the other does not. We denote these
by 650 and 650$^\prime$, respectively. In Table
\ref{tab:e6} we collect the dimension-5 contributions
for the different representations of $E(6)$.

\begin{table}
\begin{center}
\begin{tabular}{|c|c|c|c|} \hline
$E(6)$  Representations  & $\delta_{3L}$ & $\delta_{3R}$& $\delta_{3c}$  \\
\hline 
{\bf 650}  &  1/2$\sqrt2$ &  1/2$\sqrt2$& - 1/$\sqrt2$  \\
{\bf 650$^\prime$}   &  3/2$\sqrt6$ &  -3/2$\sqrt6$& 0 \\
{\bf 2430}  &  -${3/{\sqrt{26}}}$ & -${3/{\sqrt{26}}}$ &
-${3/{\sqrt{26}}}$\\
\hline
\end {tabular}
\caption{Effective contributions  ($\delta_i$)  to gauge kinetic terms from
different Higgs representations in eq. \ref{eq:dim5op} for $E(6)$
\cite{cr}.  (see eq. \ref{eq:modunif}.) Note that there are two
$SU(3)_L \otimes SU(3)_R \otimes SU(3)_c$ singlet directions
in 650 of which the first conserves D-parity while the second
does not.}
\label{tab:e6}
\end {center}
\end{table}

\subsection{No-step breaking in $E(6)$} 
This corresponds to the situation when $M_X = M_I = M_R$ and the
symmetry breaking is simply:
\begin{equation}
E(6)\stackrel{M_X}{\longrightarrow} SU(2)_L
\otimes U(1)_Y \otimes SU(3)_c.
\end{equation}
Here, eqs. \ref{eq:1lsm}, \ref{eq:2lsm}, and
\ref{eq:susy} determine the gauge coupling evolution in the
entire range. The results obtained including the dimension-5
operators in eq. \ref{eq:dim5op} are shown in Table \ref{tab:0ie6}. 

\begin{table}[hbt]
\begin{center}
\begin{tabular}{|c|c|c|c|c|}
\hline
$E(6)$&\multicolumn{2}{|c|}{Non-SUSY}&\multicolumn{2}{|c|}{SUSY} \\ \cline{2-5}
representations & $\epsilon$ & $M_X$ (GeV)& $\epsilon$ & $M_X$ (GeV)  \\
\hline 
{\bf 650} & 0.126 &8.04$\times 10^{12}$ & -0.012 &1.72$\times 10^{16}$   \\
{\bf 650$'$}  & 0.101 &4.15  $\times 10^{14}$& -0.011 &1.30  $\times 10^{16}$ \\
{\bf 2430}  & 0.000 & 3.76 $\times 10^{12}$& 0.000 & 1.25 $\times 10^{15}$  \\
\hline
\end {tabular}\\
\caption{Dimension-5 interaction strength, $\epsilon$, and the
gauge unification scale,  $M_X$, for different $\Phi_D$
representations using two-loop RG equations when $E(6)$
descends directly to the SM.}
\label{tab:0ie6}
\end {center}
\end{table}

As for the other GUT groups, though gauge unification is possible in
the non-SUSY case, the scale of unification is too low and is
ruled out by the proton decay limits. The SUSY solutions are
acceptable for $\Phi_{650}$. For $\Phi_{2430}$ the 
scale $M_X$ is too low (Note that all the $\delta_i$ are equal!)
but this can be addressed easily by changing the SUSY scale, 
$M_{SUSY}$. 
\begin{table}[hbt]
\begin{center}
\begin{tabular}{|c|c|c|c|}
\hline
$E(6)$&Symmetry&\multicolumn{2}{|c|}{Scalars contributing to RG}\\ \cline{3-4}
representation & breaking & $M_Z \rightarrow M_R$ & $M_R \rightarrow M_X$ \\
 & & Under ${\mathcal G}_{SM}$ & Under ${\mathcal
G}_{333}$   \\ \hline 

{\bf 27} & ${\mathcal G}_{SM} \rightarrow EM$ & (2,$\pm1$,1) &
($\bar{3}$,3,1)\\  
& & & \\
{\bf 27} & ${\mathcal G}_{333} \rightarrow {\mathcal G}_{SM}$ &
-  & ($\bar{3}$,3,1)\\   \hline
\end {tabular}\\
\caption{Higgs scalars for the one-step symmetry breaking of
$E(6)$ and the submultiplets contributing to RG evolution
according to the ESH.  }
\label{t:esh6s1}
\end {center}
\end{table}
\subsection{One-step breaking in $E(6)$} 
This situation corresponds to $M_I = M_R$ in eq. \ref{e6chain},
i.e.,
\begin{equation}
E(6) \stackrel{M_X}{\longrightarrow} SU(3)_L \otimes SU(3)_R \otimes
SU(3)_c \stackrel{M_R}{\longrightarrow} SM.
\end{equation}
For this case, the symmetry breaking at $M_R$ and subsequently
the one at $M_Z$ are through the $vev$s to components 
within the ($\bar{3}$,3,1) submultiplet under $SU(3)_L \otimes SU(3)_R
\otimes SU(3)_c \equiv {\mathcal G}_{333}$ which is present in a
27 of $E(6)$. According to the Extended Survival Hypothesis this
entire ($\bar{3}$,3,1) submultiplet, but for the $\phi_{SM}$
fields which are at $M_Z$,  has a mass $M_R$. Since it is
symmetric under $SU(2)_L \leftrightarrow SU(2)_R$, the evolution
of the couplings from $M_R$ to $M_X$ are controlled by the same
RG-equations for both the D-parity violating and D-parity
conserving cases. The beta-function coefficients in this case are:

{\bf From $M_R$ to $M_X$}:
\begin{equation}
{\rm \mbox{NON-SUSY:}}~~ b_{3L}=
b_{3R}=-9/2; \;\;
b_{3c}=-5;\;\;\;\;b_{ij} =\left( \begin{array}{ccc}
23 & 20 & 12 \\
20 & 23 & 12 \\
12 & 12 & 12 \\
\end{array}
\right).
\label{eq:e6UR}
\end{equation}

\begin{equation}
{\rm \mbox{SUSY:}}~~ b_{3L}= 
b_{3R}=3/2; \;\;
b_{3c}=0;\;\;\;\;b_{ij} =\left( \begin{array}{ccc}
65 & 32 & 24 \\
32 & 65 & 24 \\
24 & 24 & 48 \\
\end{array}
\right).
\label{eq:e6URS}
\end{equation}
{\bf From $M_Z$ to $M_R$}: For the RG running of the coupling
constants below $M_R$ eqs.
\ref{eq:1lsm}, \ref{eq:2lsm}, and \ref{eq:susy} are applicable
irrespective of whether D-parity is conserved or not.

{\bf Results: } The chain of $E(6)$ breaking considered in this
subsection is rather constrained. The matching formula at
$M_R$ is now:
\begin{equation}
\frac{1}{\alpha_{1Y}(M_R)} =
\frac{4}{5}\left[\frac{1}{\alpha_{3R}(M_R)} - \frac{1}{4\pi}\right] +
\frac{1}{5}\left[\frac{1}{\alpha_{3L}(M_R)} - \frac{1}{4\pi}\right] \;\; .
\label{e:match6i1}
\end{equation} 
This is a consequence of the relation $Y/2 = T_{3R} + (Y_L' + Y_R')/2$.
On the r.h.s. $T_{3R}$ and $Y_R'$ reside within the $SU(3)_R$ while $Y_L'$ is
included in $SU(3)_L$.  The two cases are: 

(a) If D-parity is not conserved then  for any chosen  $M_R$,
through eq. \ref{e:match6i1} $\alpha_{3R}(M_R)$ is fixed since
$\alpha_{2L}(M_R)$ is determined from its low energy value
through RG evolution and $1/\alpha_{3L}(M_R) = 1/\alpha_{2L}(M_R) +
1/(12 \pi)$.  The three couplings have to be further evolved to
determine $M_X$ and $\epsilon$.

(b) If D-parity is conserved at $M_R$ then in eq.
\ref{e:match6i1} $\alpha_{3R}(M_R) = \alpha_{3L}(M_R)$, with the
latter fixed by the RG evolution of ${\alpha_{2L}}$ from its low
energy value. This identifies a unique $M_R$. $M_X$ can
then be determined in terms of $\epsilon$.

We discuss these options in detail below.

When D-parity is not conserved, i.e., for $\Phi_{650'}$,
we find that the intermediate scale  at $M_R$ is rather tightly
restricted from the twin requirements that $M_X$ satisfies the proton
decay bound and is within
the upper limit set by the Planck mass as well as 
all couplings remain perturbative. It is in the ballpark of  
$10^{14}$ ($10^{16}$) GeV for the non-SUSY (SUSY) case. The
unification scale is $7.0 \times 10^{18}$ ($3.5 \times 10^{16}$)
GeV for the respective cases with $\epsilon$ almost fixed at =
-0.04 (0.02).  

When D-parity is conserved, which corresponds to $\Phi_{650}$ and
$\Phi_{2430}$, the intermediate scale $M_R$ is uniquely fixed in
both cases at the  value $1.5 \times 10^{13}$ ($1.7 \times
10^{16}$) GeV for non-SUSY (SUSY). A plot of the unification
scale $M_X$ vs. $\epsilon$ is shown in the left panel of Fig.
\ref{f:e612} for $\Phi_{650}$. For $\Phi_{2430}$ we have
$\delta_{3L} = \delta_{3R} = \delta_{3c}$ and so the dim-5
operator does not affect the unification. We find that for
non-SUSY as well as SUSY with $M_{SUSY}$ = 1 TeV  the couplings
unify at an energy beyond the Planck scale.

For both $\Phi_{650}$ and $\Phi_{650^\prime}$ the scale $M_R$ is
in the right range for the mass of the right-handed neutrinos to
drive a Type I see-saw.

\subsection{Two-step breaking in $E(6)$}\label{s:se6_2}
\noindent
The symmetry breaking steps are:
\begin{equation}
E(6) \stackrel{M_X}{\longrightarrow} SU(3)_L \otimes SU(3)_R \otimes
SU(3)_c \stackrel{M_I}{\longrightarrow} SU(2)_L \otimes U(1)_{Y'_{L}}
\otimes SU(2)_R \otimes U(1)_{Y'_{R}} \otimes SU(3)_c
\stackrel{M_R}{\longrightarrow} SM.
\end{equation}

Here, $\langle \Phi_{650}\rangle$ or  $\langle \Phi_{2430}\rangle$ 
breaks $E(6)$ to ${\mathcal G}_{333}$  which reduces to $SU(2)_L
\otimes U(1)_{Y'_{L}} \otimes SU(2)_R \otimes U(1)_{Y'_{R}}
\otimes SU(3)_c \equiv {\mathcal G}_{21213}$ when the (8,8,1)
submultiplet of a $650_H$ acquires a $vev$. The SM is reached by
assigning a $vev$ to the ($\bar{3}$,3,1) component of $27_H$. The
final step of SM symmetry breaking is accomplished through a
different component of ($\bar{3}$,3,1) (see Table \ref{t:esh6s2}).
It is seen that there is room for D-parity to be conserved or
broken during the running in the $M_R$ to $M_I$ range. But the
Higgs submultiplets which acquire masses at $M_I$ according to
the Extended Survival Hypothesis, namely, ($\bar{3}$,3,1) and
(8,8,1), are $SU(2)_L
\leftrightarrow SU(2)_R$ symmetric and so the running from $M_I$
to $M_X$ will be identical in both cases.

\begin{table}[hbt]
\begin{center}
\begin{tabular}{|c|c|c|c|c|}
\hline
$E(6)$&Symmetry&\multicolumn{3}{|c|}{Scalars contributing to RG}\\ \cline{3-5}
representation & breaking & $M_Z \rightarrow M_R$ & $M_R \rightarrow M_I$
& $M_I \rightarrow M_X$ \\
 & & Under ${\mathcal G}_{SM}$ & Under ${\mathcal
G}_{21213}$ & Under ${\mathcal
G}_{333}$   \\ \hline
{\bf 27} & ${\mathcal G}_{SM} \rightarrow EM$ & (2,$\pm1$,1) & 
(2, -$\frac{1}{2\sqrt{3}}$, 2, $\frac{1}{2\sqrt{3}}$,1) &
($\bar{3}$,3,1)\\  
& & & & \\
{\bf 27} & ${\mathcal G}_{21213} \rightarrow {\mathcal G}_{SM}$ &
-  & (1, $\frac{1}{\sqrt{3}}$, 2, $\frac{1}{2\sqrt{3}}$,1)  &
($\bar{3}$,3,1)\\ 
& & - & \{(2, $\frac{1}{2\sqrt{3}}$, 1, $\frac{1}{\sqrt{3}}$,1)\}  & \\  
& & & & \\
{\bf 650} & ${\mathcal G}_{333} \rightarrow {\mathcal G}_{21213}$ &
-  & - & (8,8,1)\\  \hline
\end {tabular}\\
\caption{Higgs scalars for the two-step symmetry breaking of
$E(6)$  and the submultiplets contributing to RG evolution
according to the ESH. The submultiplet in the braces also
contributes if D-parity is conserved.  }
\label{t:esh6s2}
\end {center}
\end{table}

It is seen from Table \ref{t:esh6s2} that in the range $M_I$ to
$M_X$ there are additional contributions from the (8,8,1) scalar
fields besides those in the one-step breaking case (Table
\ref{t:esh6s1}).  The RG evolution in the two cases is therefore
different and, as in the case of $SO(10)$,  the na\"{i}ve
expectation of the two-step results going over to the one-step one
in the limit $M_R = M_I$ does not hold.

Below we list the one- and two-loop beta-function coefficients
for gauge coupling evolution in the different stages. Notice that
in the range $M_R$ to $M_I$ there are two $U(1)$ components and
the RG evolution here has to take into account mixing and follows
the same procedure as discussed in detail for $SO(10)$ in the
previous section.

{\bf i) From $M_I$ to $M_X$}:

The fermion and scalar fields which contribute in the RG equations are:
\begin{equation}
27_F = [\bar{3},3,1]+[3,1,3]+[1,\bar{3},\bar{3}],\;\;\;\;
650_H \supset [8,8,1], \;\;\;\;
27_H \supset [\bar{3},3,1].
\end{equation}
Thus:
\begin{equation}
{\rm \mbox{NON-SUSY:}}~~ b_{3L}=7/2; \;\;
b_{3R}=7/2; \;\;
b_{3c}=-5;\;\;\;\;
b_{ij} =\left( \begin{array}{ccc}
359 & 308 & 12 \\
308 & 359 & 12 \\
12 & 12 & 12 \\
\end{array}
\right).
\label{eq:odde6UC}
\end{equation}

\begin{equation}
{\rm \mbox{SUSY:}}~~ b_{3L}=51/2; \;\;
b_{3R}=51/2; \;\;
b_{3c}=0;\;\;\;\;
b_{ij} =\left( \begin{array}{ccc}
497 & 320 & 24 \\
320 & 497 & 24 \\
24 & 24 & 48 \\
\end{array}
\right).
\label{eq:odde6UCS}
\end{equation}

{\bf iia) From $M_R$ to $M_I$ (D-parity not conserved)}:

At this stage the non-SM fermions have acquired mass and
decoupled. Taking the Extended Survival Hypothesis into
consideration, the fields that contribute in the RG equations are:
\begin{eqnarray}
27_F
&\supset&[2,-1/2\sqrt{3},1,-1/\sqrt{3},1]+[2,1/2\sqrt{3},1,0,3] +
\\ \nonumber 
& &[1,1/\sqrt{3},2,1/2\sqrt{3},1]+[1,0,2,-1/2\sqrt{3},\bar{3}], \\
\nonumber
27_H &\supset& [1,1/\sqrt{3},2,1/2\sqrt{3},1] + 
[2,-1/2\sqrt{3},2,1/2\sqrt{3},1]. \nonumber
\end{eqnarray}
This gives\footnote{The coefficients superscribed with a {\em tilde}
arise due to $U(1)$ mixing.}:
\begin{equation}
{\rm \mbox{NON-SUSY:}}~~ b_{2L}=-3; \;
b_{LL}=3; \;
b_{2R}=-17/6; \;
b_{RR}=17/6; \;
b_{3c}=-7; \;
\tilde b_{LR}= \tilde b_{RL} = 4/3,
\label{eq:1l-oddE6CR}
\end{equation}
\begin{eqnarray}
b_{ij} &=& \left( \begin{array}{ccccc}
8 & 4/3 & 3 & 4/3 & 12 \\
4 & 8/3 & 6 & 1 & 4 \\
3 & 2 & 61/6 & 3/2 & 12 \\
4 & 1 & 9/2 & 11/6 & 4 \\
9/2 & 1/2 & 9/2 & 1/2 & -26 \\
\end{array}
\right); \nonumber \\ 
\tilde b_{LR,RR} &=& 5/6;\;\;
\tilde b_{RL,LL} = 7/6;\;\;\tilde b_{2R,RL} = 1/2;\;\;\tilde
b_{2L,RL} = 1/6;\;\;\tilde b_{3c,RL} = 0;\nonumber \\
\tilde b_{RL,2R} &=& 3/2;\;\;\tilde b_{RL,2L} =1/2;\;\;
\tilde b_{RL,3c} =0;\;\;\tilde b_{RL,p} = \tilde
b_{LR,p};\;\;\tilde b_{p,LR} = \tilde b_{p,RL}.
\label{eq:2l-oddE6IR}
\end{eqnarray}

\begin{equation}
{\rm \mbox{SUSY:}}~~ b_{2L}=1; \;
b_{LL}=5; \;
b_{2R}=3/2; \;
b_{RR}=9/2; \;
b_{3c}=-3; \;
\tilde b_{LR}=\tilde b_{RL}=2,
\label{eq:1l-oddSE6CR}
\end{equation}
\begin{eqnarray}
b_{ij} &=& \left( \begin{array}{ccccc}
25 & 7/3 & 3 & 7/3 & 24 \\
7 & 13/3 & 9 & 5/3 & 8 \\
3 & 3 & 57/2 & 5/2 & 24 \\
7 & 5/3 & 15/2 & 7/2 & 8 \\
9 & 1 & 9 & 1 & 14 \\
\end{array}
\right); \nonumber \\ 
\tilde b_{LR,RR} &=& 5/3;\;\;
\tilde b_{RL,LL} = 2;\;\;\tilde b_{2R,RL} = 1;\;\;\tilde
b_{2L,RL} = 2/3;\;\;\tilde b_{3c,RL} = 0;\nonumber \\
\tilde b_{RL,2R} &=&  3;\;\;\tilde b_{RL,2L} = 2;\;\;
\tilde b_{RL,3c} =0;\;\;\tilde b_{RL,p} = \tilde
b_{LR,p};\;\;\tilde b_{p,LR} = \tilde b_{p,RL}.
\label{eq:2l-oddSE6CR}
\end{eqnarray}

{\bf iib) From $M_R$ to $M_I$ (D-parity conserved)}:

Due to D-Parity conservation the scalar sector is slightly
enlarged and  the fields contributing to the RG equations are:
\begin{eqnarray}
27_F
&\supset&[2,-1/2\sqrt{3},1,-1/\sqrt{3},1]+[2,1/2\sqrt{3},1,0,3] + \\
\nonumber & &
[1,1/\sqrt{3},2,1/2\sqrt{3},1]+[1,0,2,-1/2\sqrt{3},\bar{3}], \\
\nonumber
27_H &\supset& [1,1/\sqrt{3},2,1/2\sqrt{3},1] + 
                [2,1/2\sqrt{3},1,1/\sqrt{3},1] +
[2,-1/2\sqrt{3},2,1/2\sqrt{3},1].  \nonumber  
\end{eqnarray}
We find:
\begin{equation}
{\rm \mbox{NON-SUSY:}}~~ b_{2L}= -17/6; \;
b_{LL}= 55/18; \;
b_{2R} =-17/6; \;
b_{RR}=55/18; \;
b_{3c}=-7; \;
\tilde b_{LR}=\tilde b_{RL}=13/9,
\label{eq:1l-evenE6IR}
\end{equation}
\begin{eqnarray}
b_{ij} &=& \left( \begin{array}{ccccc}
61/6 & 3/2 & 3 & 2 & 12 \\
9/2 & 49/18 & 6 & 11/9 & 4 \\
3 & 2 & 61/6 & 3/2 & 12 \\
6 & 11/9 & 9/2 & 49/18 & 4 \\
9/2 & 1/2 & 9/2 & 1/2 & -26 \\
\end{array}
\right); \nonumber \\ 
\tilde b_{LR,RR} &=& 23/18;\;\;
\tilde b_{RL,LL} = 23/18;\;\;\tilde b_{2R,RL} = 1/2;\;\;\tilde b_{2L,LR} =
1/2;\;\;\tilde b_{3c,RL} = 0;\nonumber \\
\tilde b_{RL,2R} &=& 3/2;\;\;\tilde b_{RL,2L} = 3/2;\;\;
\tilde b_{RL,3c} =0;\;\;\tilde b_{RL,p} = \tilde
b_{LR,p};\;\;\tilde b_{p,LR} = \tilde b_{p,RL}.
\label{eq:2l-evenE6IR}
\end{eqnarray}

\begin{equation}
{\rm \mbox{SUSY:}}~~ b_{2L}=3/2; \;\;\;
b_{LL}=31/6; \;
b_{2R}=3/2; \;
b_{RR}=31/6; \;
b_{3c}=-3; \;
\tilde b_{LR}=\tilde b_{RL}=7/3,
\label{eq:1l-evenSE6CR}
\end{equation}
\begin{eqnarray}
b_{ij} &=& \left( \begin{array}{ccccc}
57/2 & 5/2 & 3 & 3 & 24 \\
15/2 & 79/18 & 9 & 17/9 & 8 \\
3 & 3 & 57/2 & 5/2 & 24 \\
9 & 17/9 & 15/2 & 79/18 & 8 \\
9 & 1 & 9 & 1 & 14 \\
\end{array}
\right); \nonumber \\ 
\tilde b_{LR,RR} &=& 19/9;\;\;
\tilde b_{RL,LL} = 19/9;\;\;\tilde b_{2R,RL} = 1;\;\;\tilde
b_{2L,LR} = 1;\;\;\tilde b_{3c,RL} = 0;\nonumber \\
\tilde b_{RL,2R} &=&  3;\;\;\tilde b_{RL,2L} = 3;\;\;
\tilde b_{RL,3c} =0;\;\;\tilde b_{RL,p} = \tilde
b_{LR,p};\;\;\tilde b_{p,LR} = \tilde b_{p,RL}.
\end{eqnarray}

{\bf From $M_Z$ to $M_R$}: For the RG running of the coupling
constants below $M_R$ eqs.
\ref{eq:1lsm}, \ref{eq:2lsm}, and \ref{eq:susy} are applicable
irrespective of whether D-parity is conserved or not.

\begin{center}
\begin{figure}[thb]
\hskip .50cm
\includegraphics[width=5.5cm,height=4.50cm,angle=0]{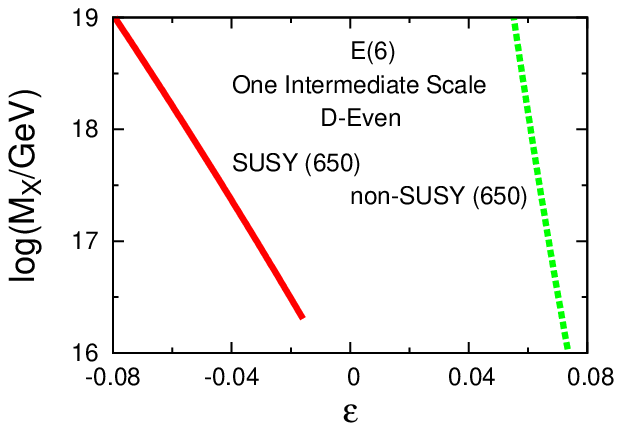}
\hskip 0.1cm
\includegraphics[width=5.2cm,height=4.5cm,angle=0]{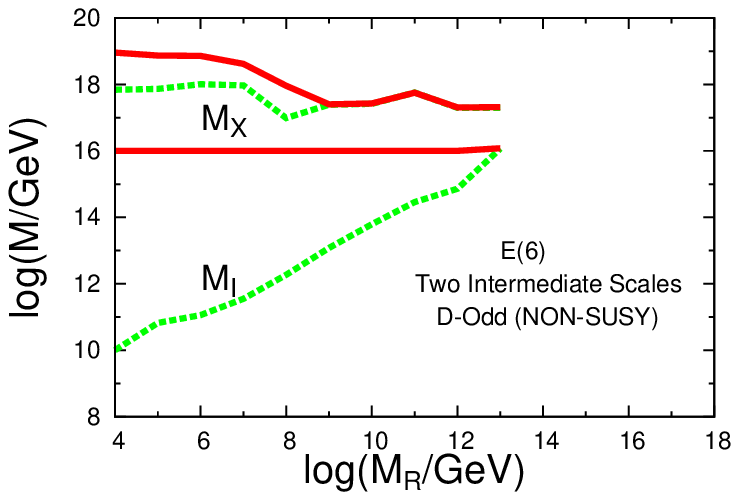}
\hskip 0.1cm
\includegraphics[width=4.9cm,height=4.5cm,angle=0]{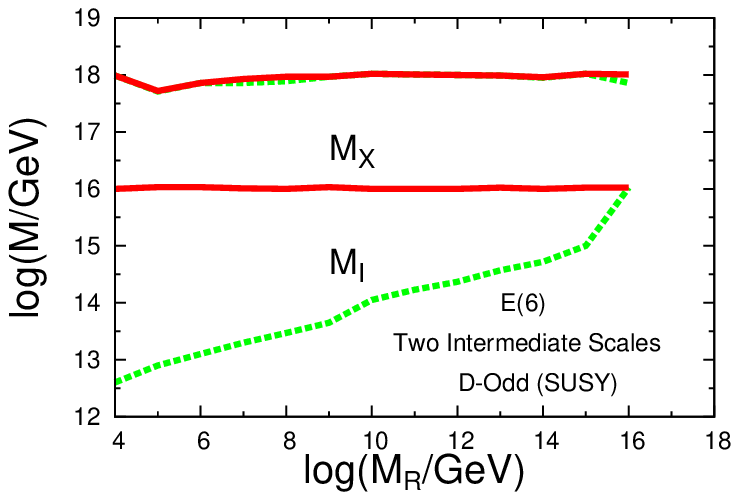}
\caption{\sf \small {Left panel: $M_X$ as a function of
$\epsilon$ for one-step breaking of $E(6)$ in the D-parity conserving
case for $\Phi_{650}$.  The  green pale broken (red
dark solid) line corresponds to non-SUSY (SUSY). 
Centre and Right panels: The allowed ranges of $M_X$ (red dark
solid) and $M_I$ (green pale broken) vs. $M_R$ for the non-SUSY
(centre) and SUSY (right) cases for $E(6)$ breaking through two
intermediate steps when D-parity is not conserved.  Note that the
upper limits for $M_X$ and $M_I$ are almost identical.}}
\label{f:e612}
\end{figure}
\end{center}

{\bf Results: }  
When $E(6)$ breaks to the SM through two intermediate steps, 
at $M_R$ one must set:
\begin{equation}
\frac{1}{\alpha_{1Y}(M_R)} =
\frac{3}{5}\left[\frac{1}{\alpha_{2R}(M_R)} - \frac{1}{6\pi}\right]+
4 \pi ~P \; (G \; G^T)^{-1}  P^T \;.
\label{e:match6i2}
\end{equation} 
where $P = (\sqrt{\frac{1}{5}} \;\; \sqrt{\frac{1}{5}}) $,
which follows from $Y/2 = T_{3R} + (Y_L' + Y_R')/2$.

When the initial symmetry breaking of $E(6)$ is through the
$\Phi_{650'}$, D-parity is not conserved. It might seem that
there is some flexibility here and at $M_R$ one can choose
$g_{Y'_R Y'_R}$, $g_{Y'_R Y'_L}$, and $g_{2R}$ independently, determining
$g_{Y'_L Y'_L}$ from eq. \ref{e:match6i2}. In fact, there is a
rather severe constraint that $\alpha_{Y'_R}$ and $\alpha_{2R}$
must meet at $M_I$ and at precisely the same scale
$\alpha_{Y'_L}$ must equal $\alpha_{2L}$.  In 
Fig. \ref{f:e612} we show the allowed  range of
the intermediate scale $M_I$ and the unification scale $M_X$ as a
function of $M_R$. Note that for both cases these scales are on
the high side. The scale of the second stage of symmetry
breaking, $M_R$, is permitted to be as low as $10^4$ GeV for the
non-SUSY as well as the SUSY case. It determines the mass scale
of a $Z'$ boson and  may offer room for experimental probing at
the LHC. The right-handed charged weak bosons are at $M_C$ and
hence beyond reach. $\epsilon$ is bounded in the range $0 \leq
|\epsilon | \leq 0.16$.

When the first stage of symmetry breaking is driven through the
$\Phi_{650}$, D-parity is preserved. This implies that
$\alpha_{2R}(M_R) = \alpha_{2L}(M_R)$ and is fixed by the RG
evolution of $g_{2L}$ from $M_Z$. Also at $M_R$, $g_{Y'_L Y'_L} =
g_{Y'_R Y'_R}$ and one can choose $g_{Y'_L Y'_R} = g_{Y'_R Y'_L}
= 0$, so all couplings are determined once $M_R$ is chosen.
Requiring that the constraint on $M_X$ from proton decay be
satisfied along with perturbativity, we find that there is a very
limited range of allowed solutions with $10^{11}$ GeV $\leq  M_R
\leq ~10^{13}$ GeV (non-SUSY case) and  $10^{15}$ GeV $\leq  M_R
\leq ~10^{16}$ GeV (SUSY case). $M_I$ and $M_X$ are always close
together around $10^{16-17}$ GeV. For these solutions
$0 \leq |\epsilon | \leq 0.14$.

The case of $\Phi_{2430}$ is not distinguishable from the
situation of no dim-5 operators at all since here $\delta_1 =
\delta_2 = \delta_3$.

\section{Summary and Discussions}

In this paper we have examined the GUT-symmetry breaking
consequences of dim-5 operators which can arise from quantum
gravity or string compactification leading to a correction to the gauge
kinetic term. When the GUT symmetry is broken their effect is to
modify gauge coupling unification to the relation 
$g_{i}^{2}(M_{X})(1+\epsilon\delta_i)= g_{U}^2$ (eq.
\ref{eq:modunif}). The relevant group theoretic  factors
$\delta_i$ were exhaustively calculated in
\cite{cr}. Here we have focussed on the implications for
grand unification and intermediate energy scales, both for single and
multi-step breaking and also for non-supersymmetric as well as
supersymmetric theories. We have required all coupling constants
to remain perturbative in the entire energy range and that the
bound on the GUT scale from non-observation of proton decay be
respected.  We have remarked on $n-\bar{n}$ oscillations and
see-saw light neutrino mass  implications in passing.

For multi-step symmetry breaking cases we have utilised the
Extended Survival Hypothesis to decide which scalar
submultiplet gets mass at which scale. When there are two $U(1)$
factors at some intermediate stage, we consider the effect of
their mixing. 

For $SU(5)$ we show that even after the inclusion of the effect
of dim-5 operators the non-SUSY version cannot be rescued from
the proton decay limit impasse while the SUSY version works fine
not just when the initial GUT breaking is through the usual
$\Phi_{24}$ but also by $\Phi_{75}$ and $\Phi_{200}$.

For $SO(10)$ we consider the direct breaking to the SM as well as
multi-step breaking {\em via} the Pati-Salam ${\mathcal G}_{224}$
route. For the former case, the conclusions are pretty much the
same as that for $SU(5)$. For the latter alternative, the
spontaneous symmetry breaking can be achieved through
$\Phi_{54}$, $\Phi_{210}$, and $\Phi_{770}$. We classify the
solutions according to whether (a) they conserve D-parity 
($\Phi_{54}$ and $\Phi_{770}$) or (b) not ($\Phi_{210}$).
(b) turns out to be phenomenologically more interesting. 
If there is one intermediate
scale then in (b) this can be as low as $10^3$ GeV with a
plethora of observable consequences including charged and neutral
gauge bosons and a possibility of observable $n-\bar{n}$
oscillations. For (a) this scale is very high: $10^{13}$ GeV or
more.  This is also the energy at which $\nu_R$ develops a mass
and so it could conveniently generate light neutrino masses with
${\cal O}(1)$ Yukawa couplings. In the case of two intermediate
scales, for both (a) and (b) one can have one of them
as low as 1 TeV where  a neutral gauge boson is expected. The
other scale can be $10^{6.5}$ GeV or higher for (b) and $10^{13}$
GeV or more for (a).

For $E(6)$ the GUT symmetry breaking can be achieved through two
possible $vev$s for the 650-dimensional Higgs scalar multiplet,
which we call $\Phi_{650}$ and $\Phi_{650'}$ as well as through a
$\Phi_{2430}$.   For the direct breaking to the SM the results
are again as in the case of other GUT groups, namely, the
non-SUSY case is disfavoured and the SUSY option is consistent
with all requirements. For multistep breaking we consider the
${\mathcal G}_{333}$ route. Here the solutions that we obtain
with $\Phi_{650}$ and  $\Phi_{650'}$ all have options with one
intermediate scale as low as $10^{4}$ GeV or higher. For
$\Phi_{2430}$, $\delta_{3L} = \delta_{3R} =
\delta_{3c}$ and the situation remains identical to the usual case
but for a scaling of the unified coupling. 

A general remark about two-step and one-step breaking 
is that the additional scalar fields which
drive the  symmetry breaking at $M_C$ for $SO(10)$ ($M_I$ for
$E(6)$) in the former case contribute in the RG evolution in the
stage $M_C \rightarrow M_X$  ($M_I \rightarrow M_X$) over and
above whatever is present in the one-step breaking case. Due to
this, the simple-minded expectation of the two-step case going
over to the one-step case in the limit of $M_R = M_C$ for
$SO(10)$ and $M_R = M_I$ for $E(6)$ is not valid.

Finally, we would like to compare our results with some
of the earlier analyses of GUT symmetry breaking with
intermediate scales, albeit without dim-5 operators. Multistep
symmetry breaking of $SO(10)$ has been looked at, for
example, in
\cite{Berto} and \cite{oldso10}. For the chain of eq.
\ref{so102s}, i.e., {\em via} ${\mathcal G}_{224}$ and ${\mathcal
G}_{2131}$ no acceptable solutions were found in the non-SUSY
case in \cite{Berto} while in \cite{oldso10} solutions with $M_R$
in the range $10^5 - 10^7$ GeV were presented\footnote{In
\cite{oldso10}, the $U(1)$ mixing contributions to the RG
equations at one loop were not included. We also have 
differences with a few beta-function coefficients calculated 
in this paper.}.  Here we obtained a
wider span of $10^4 - 10^{16}$ GeV. For the one-step
symmetry breaking of non-SUSY $SO(10)$ the scale $M_C$ was found
in \cite{oldso10} to be in the $10^5 - 10^7$ GeV range whereas
with the inclusion of dim-5 operators we have shown that this
scale is in the phenomenologically attractive $10^3 - 10^{10}$
GeV region. 

$SU(5)$ GUT with the inclusion of the dim-5 operator from
$\Phi_{24}$ has been examined in \cite{Vayon}. The results for
non-SUSY as well as the SUSY cases are in agreement with the ones
in sec. \ref{s:su5}. Our analysis also covers $\Phi_{75}$ and
$\Phi_{200}$ of $SU(5)$. For $SO(10)$ the effect of $\Phi_{54}$ and
$\Phi_{210}$ has been considered in \cite{Parida} using the
one-loop RG equations. They also noted, like us, that for $\Phi_{54}$, when
D-parity is conserved, the scale $M_C$ is uniquely fixed by the
measured $\sin^2\theta_W$ and the $M_X$ they obtain is in
agreement with our results. For the D-parity non-conserving case
of $\Phi_{210}$ they find a range of $M_C$ and $M_X$ similar to
what is depicted in Fig. \ref{f:s1ns}.

As regards $E(6)$, we could not trace any earlier published
analysis in the descent to the SM through the ${\mathcal
G}_{333}$ chain \cite{olde6rs}. The attention has invariably
focussed on $E(6)$ breaking through an intermediate
$SO(10) \times U(1)$ \cite{olde6}. 

It can be hoped that further refinement in the determination of
the low energy gauge couplings, proton decay tests, and
explorations of $n-\bar{n}$ oscillation  will enable us to
extract signals of physics that lies beyond the grand unification
scale.


\vskip 20pt

{\bf Acknowledgements} This research has been supported by funds
from the DAE XIth Plan `Neutrino Physics' and `RECAPP' projects
at HRI.  The HRI cluster computational facility has been utilised
for the numerical work.

\end{document}